\pdfoutput=1
\documentclass[runningheads]{llncs}
\usepackage[flushleft]{threeparttable}

\usepackage{url}
\usepackage{amsmath}
\usepackage{amsfonts,bm}
\usepackage{subfig}
\usepackage{booktabs}
\usepackage{graphicx}
\usepackage{todonotes}
\usepackage[colorlinks=true,urlcolor= blue]{hyperref}%
\usepackage{bbding}
\usepackage{multirow}
\usepackage{textcomp}
\makeatletter
\newcommand{\printfnsymbol}[1]{%
  \textsuperscript{\@fnsymbol{#1}}%
}
\makeatother

\newcommand{\mat}[1]{{\boldsymbol{{#1}}}} % matrix
\newcommand{\mcl}[1]{{\mathcal{{#1}}}} % mathcal

\providecommand{\mx}{\mat{x}}

\providecommand{\my}{\mat{y}}

\providecommand{\mc}{\mat{c}}
\providecommand{\ms}{\mat{s}}

\begin{document}
%Suggested titles listed here
\title{Unsupervised Multi-modal Style Transfer for Cardiac MR Segmentation}

\author{Chen Chen\inst{1}\thanks{Equal contribution} (\Envelope), Cheng Ouyang\inst{1}\printfnsymbol{1} (\Envelope), Giacomo Tarroni\inst{1}, Jo Schlemper\inst{1}, \\ Huaqi Qiu\inst{1}, Wenjia Bai\inst{2,3}, Daniel Rueckert\inst{1}}

\institute{Biomedical Image Analysis Group, Imperial College London, London, UK\\
\and Data Science Institute, Imperial College London, London, UK
\and Department of Medicine, Imperial College London, London, UK% %
\\
\email{\{chen.chen15, c.ouyang\}@imperial.ac.uk}}

%
\authorrunning{Chen., Ouyang. et al.}
\maketitle              % typeset the header of the contribution
\begin{abstract}
In this work, we present a fully automatic method to segment cardiac structures from late-gadolinium enhanced (LGE) images without using labelled LGE data for training, but instead by transferring the anatomical knowledge and features learned on annotated balanced steady-state free precession (bSSFP) images, which are easier to acquire. Our framework mainly consists of two neural networks:  a multi-modal image translation network for style transfer and a cascaded segmentation network for image segmentation. The multi-modal image translation network generates realistic and diverse synthetic LGE images conditioned on a single annotated bSSFP image, forming a synthetic LGE training set. This set is then utilized to fine-tune the segmentation network pre-trained on labelled bSSFP images, achieving the goal of unsupervised LGE image segmentation. In particular, the proposed cascaded segmentation network is able to produce accurate segmentation by taking both shape prior and image appearance into account, achieving an average Dice score of 0.92 for the left ventricle, 0.83 for the myocardium, and 0.88 for the right ventricle on the test set.
\end{abstract}
\section{Introduction}
Cardiac segmentation from late-gadolinium enhanced (LGE) cardiac magnetic resonance (CMR) which highlights myocardial infarcted tissue is of great clinical importance, enabling quantitative measurements useful for treatment planning and patient management. To this end, the segmentation of the myocardium is an important first step for myocardial infarction analysis.

\begin{figure}[!ht]
    \centering
    \includegraphics[width=0.8\textwidth]{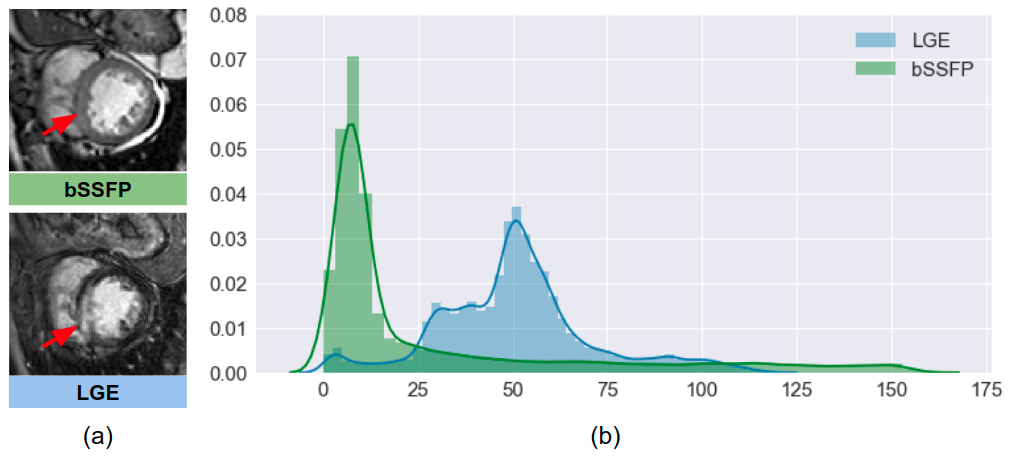}
    \caption{The differences of image appearance (a) and intensity distributions (b) in the cardiac region (the union of LV, MYO, RV) between LGE images and bSSFP images}
    \label{fig:intensity distribution}
\end{figure}

Since manual segmentation is tedious and likely to suffer from inter-observer variability, it is of great interest to develop an accurate automated segmentation method. However, this is a  challenging task due to the fact that 1) the infarcted myocardium presents an enhanced and heterogeneous intensity distribution different from the normal myocardium region and 2) the border between infarcted myocardium and blood pool appears blurry and ambiguous~\cite{zhuang2016multivariate}. While the borders of the myocardium can be difficult to delineate on LGE images, they are clear and easy to identify on the balanced steady-state free precession (bSSFP) CMR images, which have high signal-to-noise ratio and whose contrast is less sensitive to pathology (see red arrows in Fig.~\ref{fig:intensity distribution} (a)). Conventional methods~\cite{lu2013automatic,tao2015automated} use the segmentation result from the bSSFP CMR of the same patient as prior knowledge to assist the segmentation on LGE CMR images. These methods generally require accurate registration between the bSSFP and LGE images, which can be challenging as the imaging field-of-view (FOV), image contrast and resolution between the two acquisitions can vary significantly \cite{zhuang2016multivariate,Zhuang2018MultivariateMM}. Fig.~\ref{fig:intensity distribution} (b) visualizes the discrepancy between the intensity distributions of the two imaging modalities in the cardiac structures (specifically, left ventricle (LV), myocardium (MYO), and right ventricle (RV)).

Most recently, a deep neural network-based method has been proposed to segment the three cardiac structures directly from LGE images~\cite{yue2019cardiac}, reporting superior performance. However, this supervised segmentation method requires a large amount of labelled LGE data. Because of the heterogeneous intensity distribution of the myocardium in LGE images and the scarcity of experienced image analysts, it is difficult to perform accurate manual segmentations on LGE images and collect a large training set, compared to that on bSSFP images.

In this paper, we present a fully automatic framework that addresses the above mentioned issues by training a segmentation model without using manual annotations on LGE images. This is achieved by transferring the anatomical knowledge and features learned on annotated bSSFP images, which are easier to acquire. Our framework mainly consists of two neural networks:
\begin{itemize}
    \item A multi-modal image translation network: this network is used for translating annotated bSSFP images into LGE images through style transfer. Of note, the network is trained in an unsupervised fashion where the training bSSFP images and LGE images are \textbf{unpaired}. In addition, unlike common one-to-one translation networks, this network allows the generation of \textbf{multiple} synthetic LGE images conditioned on a single bSSFP image;
    \item A cascaded segmentation network for LGE images consisting of two U-net~\cite{ronneberger2015u} models (Cascaded U-net): this network is first trained using the labelled bSSFP images and then fine-tuned using the synthetic LGE data generated by the image translation network. 
    % Specifically, this cascaded network consists of two U-nets where the second U-net utilizes the predicted probabilistic maps produced by the first U-net to assist the segmentation. By combining the object \textbf{shape} information from the probabilistic maps and image \textbf{appearance} information from the input image, the network is able to produce precise segmentation.
\end{itemize}

The main contributions of our work are the following: 1) we employ a translation network that can generate \textbf{realistic} and \textbf{diverse} synthetic LGE images given a single bSSFP image. This network enables generative model-based data augmentation for unsupervised domain adaptation, which not only closes the domain gap between the two modalities, but also improves the generalization properties of the following segmentation network by increasing data variety; 2) we demonstrate that the proposed two-stage cascaded network, which takes both anatomical \textbf{shape} information and image \textbf{appearance} information into account, produces accurate segmentation on LGE images, greatly outperforming baseline methods; 3) the proposed framework can be easily extended to other unsupervised cross-modality domain adaptation applications where labels of one modality are not available.
% ; 4) To the best of our knowledge, this fully-automatic framework is the first neural network-based method for \textbf{unsupervised} LGE image segmentation. 

\section{Methodology}
The proposed method aims at learning an LGE image segmentation model using labelled bSSFP $\{ (\mx_b, \my_b) \}$ and unlabelled LGE $\{ \mx_l \}$ only. 
% Here subscripts $b$ pr $l$ stand for bSSFP or LGE domain.%
Specifically, the proposed method is a two-stage framework. In the first stage, an unsupervised \textbf{image translation} network is trained to translate each bSSFP image $\mx_b$ into multiple instances of LGE-like images, noted as $\{\mx_{bl}\}$. In the second stage, these LGE-stylized bSSFP images are used together with their original labels $\{(\mx_{bl}, \my_b)\}$ to adapt an \textbf{image segmentation} network pre-trained on labelled bSSFP images to segment LGE images.
\subsection{Image Translation}
 We employ the state-of-the-art multi-modal unsupervised image-to-image translation network (MUNIT)~\cite{huang2018multimodal} as our multi-modal image translator. Let $\{\mx_{l}\}$ and $\{\mx_{b}\}$ denote unpaired images from the two different imaging modalities (domains): LGE and bSSFP, given an image drawn from one domain as input, the network is able to change the appearance (i.e. image style) of the image to that of the other domain while preserving the underlying anatomical structure~\cite{Qin_2019_IPMI}. This is achieved by learning disentangled image representations.

\begin{figure}[!ht]
\centering
\includegraphics[width=0.8\textwidth]{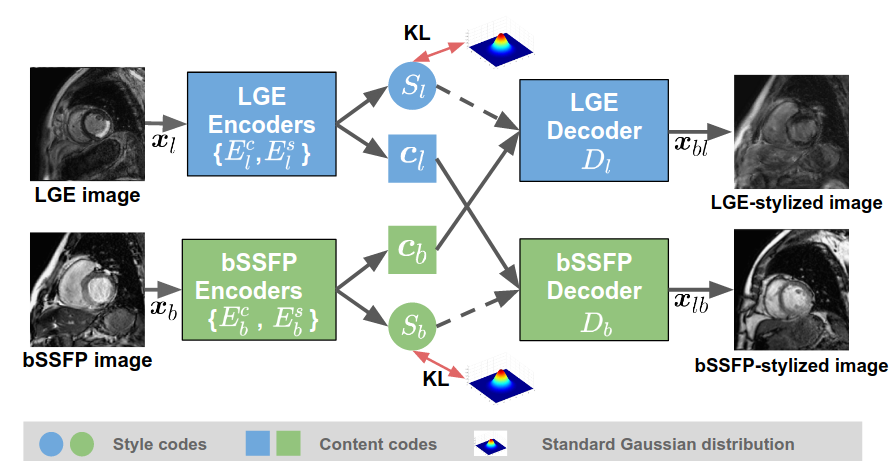}
\caption{\textbf{Overview of the multi-modal image translation network.} The network employs the structure of MUNIT~\cite{huang2018multimodal}, which consists of two encoder-decoder pairs for the two domains: bSSFP and LGE, respectively.}
\label{fig:munit}
\end{figure}
As shown in Fig. \ref{fig:munit}, each image $\mx$ is disentangled into (a) a domain-invariant content code $\boldsymbol{c}$: $\boldsymbol{c}=E^c{(\mx)}$ and (b) a domain-specific style code $\boldsymbol{s}$:  $\boldsymbol{s}=E^s{(\mx)}$
using the content encoder $E^c$ and the style encoder $E^s$ relative to its domain where the content code
captures the anatomical structure and the style code carries the information for rendering the structure which is determined by the imaging modality. The image-to-image translation from one domain to the other is achieved by swapping latent codes in two domains. For example, translating a bSSFP image $\boldsymbol{x_b}$ to be stylized as LGE, is achieved by feeding the content code  $\boldsymbol{c_b}$ for the bSSFP image and the style code $\boldsymbol{s_l}$ into the LGE decoder $D_l$: $\mx_{bl} =D_{l}(\mc_{b},\ms_{l})$. 

Of note, during training, each style encoder is trained to embed images into a latent space that matches the standard Gaussian distribution $\mcl{N}(0, I)$, minimizing the Kullback-Leibler (KL) divergence between the two. This allows to generate an arbitrary number of synthetic LGE images given a single bSSFP image during inference, by repeatedly sampling the style code from the prior distribution $\mcl{N}(0, I)$. Of note, although this prior distribution is unimodal, the distribution of translated images in the output space is multi-modal thanks to the nonlinearity of the decoder\cite{huang2018multimodal}. We apply this translation network to translate annotated bSSFP images, resulting in a synthetic labelled LGE dataset, which will then be used to finetune a segmentation network.    
For more details about training the translation network, readers are referred to the original work by Huang \textit{et al.} \cite{huang2018multimodal}.

\subsection{Image Segmentation}
\begin{figure}[!ht]
    \centering
    \includegraphics[width=\textwidth]{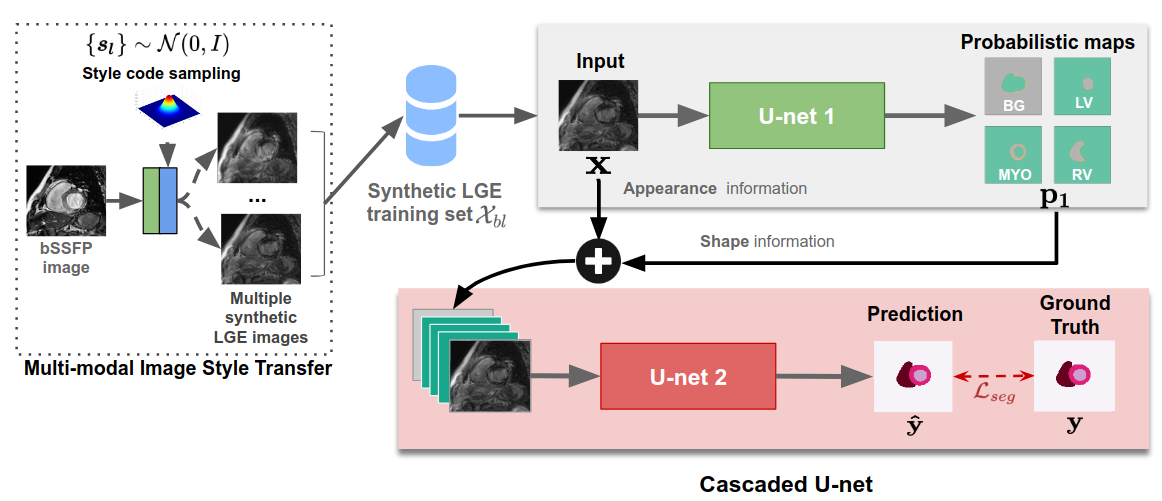}
    \caption{\textbf{Overview of the two-stage cascaded segmentation network.} The architecture of each U-net is the same as the one of the vanilla U-net, except for two main differences: (1) batch normalization is applied after each convolutional layer; (2) a dropout layer (dropout rate=0.1) is applied after each concatenation operation in the network's expanding path to encourage model generalizability. Of note, in this diagram, we simplify the training procedure by omitting the pre-training procedure using labelled bSSFP images.}
    \label{fig:cascaded unet}
\end{figure}
Let $\boldsymbol{x}_{l}$ be an observed LGE image, the aim of the segmentation task is to estimate label maps $\boldsymbol{y}_{l}$ having
observed $\boldsymbol{x}_{l}$ by modeling the posterior $p(\boldsymbol{y}_{l}|\boldsymbol{x}_{l})$. Inspired by curriculum learning~\cite{Bengio2009} and transfer learning, we first train a segmentation network using annotated bSSFP images (source domain; easy examples) and then fine-tune it to segment LGE images (target domain; hard examples). Since labelled LGE images $\{(\mx_{l}, \my_{l})\}$ are not available for finetuning, we use a synthetic dataset  $\mathcal{X}_{bl}: \{ (\mx_{bl},\my_{b})\}_{1..N}$ generated by the aforementioned multi-modal image translator. Ideally, the posterior modelled by the network $p(\boldsymbol{y}_{b}|\boldsymbol{x}_{bl})$ matches $p(\boldsymbol{y}_{l}|\boldsymbol{x}_{l})$ when image space and label space are shared. For simplicity, we use $\boldsymbol{x}$ and $\boldsymbol{y}$ to denote an image and its corresponding label map from the synthetic dataset in the following paragraphs.

The segmentation network is a two-stage cascaded network which consists of two U-nets \cite{ronneberger2015u}, see Fig.~\ref{fig:cascaded unet}. Specifically, given an image $\boldsymbol{x}$ as input, the first U-net (U-net 1) aims at predicting four-class pixel-wise probabilistic maps $\boldsymbol{p_1}=f_{\text{U-net}}^{1}(\boldsymbol{x}; \theta)$ for the three cardiac structures (i.e. LV, MYO, RV) and the background class (BG). Inspired by the auto-context architecture \cite{Tu2010}, we combine these learned probabilistic maps $\boldsymbol{p_1}$ from the first network with the raw image $\boldsymbol{x}$ to form a 5-channel input to train the second U-net (U-net 2) for fine-grained segmentation: $\boldsymbol{{p_2}} ={{f_\text{U-net}^2}}(\boldsymbol{x},\boldsymbol{p_1};\phi)$. By combining the appearance information from the image $\mx$ with the shape prior information from the initial segmentation $\boldsymbol{p_1}$ as input, the cascaded network has the potential to produce more precise and robust segmentations even in the presence of unclear boundaries for the different cardiac structures. 

To train the network, we use a \textbf{composite segmentation loss} function $\mathcal{L}_{seg}$ which consists of two loss terms:
$    \mathcal{L}_{seg} = \mathcal{L}_{wce}+\lambda \mathcal{L}_{edge}.
$ The first term $\mathcal{L}_{wce}$ is a weighted cross entropy loss:  
$
    \mathcal{L}_{wce}=-\sum_{m} \omega^{m} \boldsymbol{y}^{m} \log \left(\boldsymbol{p}^{m}\right) 
$
where $w^{m}$ denotes the weight for class $m$ and $\boldsymbol{p}^{m}$ is the corresponding predicted probability map. We set the weight for myocardium $\omega^{MYO}$ to be higher than the weights for the other three classes to address class imbalance problem since there is a lower
percentage of pixels that corresponds
to the myocardium class in CMR images.  
The second term $\mathcal{L}_{edge}$ is an edge-based loss which penalizes the disagreement on the contours of the cardiac structures. Specifically, we apply two 2D $3 \times 3$ Sobel filters \cite{sobel} $S_k$ (k=1,2) to the soft prediction maps $\boldsymbol{p}$ as well as the one-hot heatmaps $\boldsymbol{y}$  of the ground truth to extract edge information along horizontal and vertical directions. 
% \begin{equation}
%   S_{1}=\begin{vmatrix}
%  -1 \;\;&0 \;\; & 1\;\; \\ 
% -2 \;\;& 0\;\;& 2\;\; \\ 
%  -1 \;\;& 0 \;\;& 1 \;\;
% \end{vmatrix},
%  S_{2}=\begin{vmatrix}
%  -1& -2& -1 \\ 
%  0 & 0 & 0 \\ 
%  1 & 2 & 1 
% \end{vmatrix}  
% \end{equation}
The edge loss is then computed by calculating the $l_{2}$ distance between the predicted edge maps and the ground truth edge maps:
$ \mathcal{L}_{edge}=\sum_{m, m \neq BG} \sum_{k=1,2} \left\|f_{S_{k}}({\boldsymbol{p}^m})-f_{S_{k}}(\boldsymbol{y}^{m})\right\|_{2}
$, 
where $f_{S_{k}}(\boldsymbol{p}^{m})$ is the edge map extracted by applying the sobel filter $S_{k}$ to the predicted probabilistic map $\boldsymbol{p}^{m}$ for foreground class $m$.

By using the edge loss together with the weighted cross entropy for optimization, the network is encouraged to focus more on the contours of the three structures and the myocardium, which are usually more difficult to delineate. In our experiments, we set $\lambda$ = 0.5 to balance the contribution of the two losses.

\subsection{Post-processing}
At inference time, each slice from a previously unseen LGE stack is fed to the cascaded network to get the probabilistic maps for the four classes. 
% To achieve better predictive performance, we employ ensemble learning by applying the Monte Carlo dropout~\cite{gal2016dropout}, performing $T$ stochastic forward passes through the network and averaging the results. In this procedure, each dropout layer in the second U-net is activated with a probability of 0.1. %
Dense conditional random field (CRF)~\cite{krahenbuhl2011efficient} is then applied to refine the 2D predicted segmentation mask slice by slice. After that, 3D morphological dilation and erosion operations are applied to the whole segmentation stack to further improve the global smoothness. In particular, we perform the operations in a hierarchical order: first we apply them to the binary map covering all the three structures, then to the MYO and the LV labels, separately.

\section{Experiments and Results}
\label{sec:implementation}
\subsection{Data} The framework was trained and evaluated on the Multi-sequence Cardiac MR Segmentation Challenge (MS-CMRSeg 2019) dataset\footnote{\url{https://zmiclab.github.io/mscmrseg19/}}.
We used a subset of 40 bSSFP and 40 LGE images to train the image translation network. Then, we created a synthetic dataset by applying the learned translation network to 30 labelled bSSFP images. Specifically, for each bSSFP image, we randomly sampled the style code from $\mcl{N}(0, I)$ five times, resulting in a set of 150 synthetic LGE images in total. This synthetic dataset and the original 30 bSSFP images with corresponding labels formed the training set for the segmentation network. Exemplar results of these synthetic LGE images are provided in the supplemental material. For validation, we used a subset of 5 annotated LGE images provided by the challenge organizers.

\subsection{Implementation Details}
\noindent\textbf{Image preprocessing.} To deal with the different image size and heterogeneous pixel spacing between different imaging modalities, all images were resampled to a pixel spacing of $1.25~mm \times 1.25~mm$ and then cropped to $192 \times 192 $ pixels, with the heart roughly at the center of each image. This spatial normalization would reduce the computational cost and task complexity in the following training procedure of image translation and segmentation, making the networks focus on the relevant regions. To identify the heart, we trained a localization network based on U-net using the 30 annotated bSSFP images in the training set to produce rough segmentations for the three structures. The localization network employs instance normalization layers which perform style normalization~\cite{huang2017arbitrary}, encouraging the network invariance to image style changes (e.g. image contrast). As a result, the network is able to produce coarse masks localizing the heart on all bSSFP images and most LGE images even though it was trained on bSSFP images only. In case that this network might fail to locate the heart on certain LGE slices, we summed the segmentation masks across slices in each volume and then cropped them according to the center of the aggregated mask. After cropping, each image was intensity normalized.
\newline

\noindent\textbf{Network training.} (1) For the image translation network, we used the official implementation\footnote{\url{https://github.com/NVlabs/MUNIT}} of \cite{huang2018multimodal}. Network configuration and hyper-parameters were kept the same as in \cite{huang2018multimodal} except the input and output images are 2D, single-channel. It was trained for 20k iterations with a batch size of 1. (2) For the segmentation network, we first trained the first U-net with the labelled bSSFP images and then fine-tuned it with synthetic LGE images. This procedure was replicated to train the second U-net with the parameters of the first U-net being fixed. Both networks were optimized using the composite loss $\mathcal{L}_{seg}$ where adam was used for stochastic gradient descent. The learning rate was initially set to 0.001 and was then decreased to $1 \times 10^{-5}$ for fine-tuning. The weights for BG, LV, MYO, and RV in $\mathcal{L}_{wce}$ were empirically set to $0.2:0.25:0.3:0.25$. 
During training, we applied data augmentation on the fly. Specifically, elastic deformations, random scaling and random rotations as well as gamma augmentation~\cite{chen_2019} were used. The algorithm was implemented using python and PyTorch and was trained for 1000 epochs in total on an NVIDIA$^{\tiny{\text{\textregistered}}}$
Tesla P40 GPU.
\subsection{Results}
To evaluate the accuracy of segmentation results, the Dice metric and the average surface distance (ASD)
between the automatic segmentation and the corresponding manual segmentation for each volume were calculated.

\begin{figure}[!ht]
    \centering
    \includegraphics[width=\textwidth]{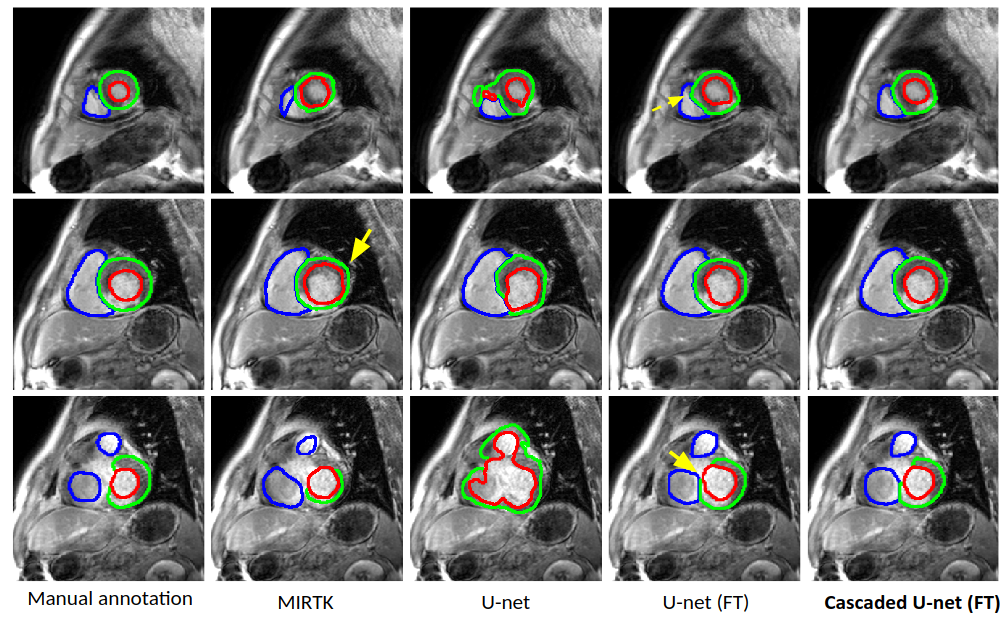}
    \caption{\textbf{Segmentation results for the proposed Cascaded U-net and the baseline approaches}. Our proposed method (the right-most column) produces more anatomically plausible segmentation results on the images, greatly outperforming the baseline methods, especially in the challenging cases: the apical (the top row) and the basal slices (the bottom row).}
    \label{fig:segmentation_result}
\end{figure}

We compare the proposed method with two baseline methods: (1) a registration-based method and (2) a single U-net. Specifically, for the registration-based method, each LGE segmentation result was obtained by directly registering the corresponding bSSFP labels to the LGE image using MIRTK toolkit \footnote{\url{https://mirtk.github.io/}} for ease of comparison. The transformation matrix was learned by applying mutual information-based registration (Rigid+Affine+FFD) between the two images. For U-net, we trained it with two settings: a) \textbf{U-net}: trained on labelled bSSFP images only; b) \textbf{U-net with fine-tuning (FT)}: trained on labelled bSSFP images and then fine-tuned using the synthetic LGE data, which is the same training procedure of the proposed method. Quantitative and qualitative results are shown in Table~\ref{tab: result table} and Fig.~\ref{fig:segmentation_result}.

While the registration-based method (MIRTK) outperforms the U-net (see row 1 and row 2 in Table \ref{tab: result table}), it still fails to produce accurate segmentation on the myocardium (see the \textcolor{red}{red number} in row 1), indicating the limitation of this registration-based method. However, by contrast, neural network-based methods (row 3-5) fine-tuned using the \emph{synthetic LGE dataset} significantly improves the segmentation accuracy, increasing the Dice score for MYO by $\sim 15\%$. This improvement demonstrates the learned translation network is capable of generating realistic LGE images while preserving the domain-invariant structural information that is informative to optimize the segmentation network. In particular, compared to U-net (FT), the proposed \textbf{Cascaded U-net} (FT) achieves more accurate segmentation performance with improvement in terms of both Dice and ASD (see \textcolor[HTML]{3531FF}{blue numbers}). The model even produces robust segmentation results on the challenging apical and basal slices (please see the last column in Fig. \ref{fig:segmentation_result}). This demonstrates the benefit of integrating the high-level shape knowledge and low-level image appearance to guide the segmentation procedure. In addition, the proposed post-processing further refines the segmentation results through smoothing, reducing the average ASD from 1.37 to 1.26 (see the last row in Table~\ref{tab: result table}).   

\begin{table}[!ht]
\centering
\caption{\textbf{Dice scores and ASD (mm) of the proposed segmentation method (Cascaded U-net) and baseline methods on the validation set.} \small{\textcolor[HTML]{3531FF}{Blue numbers} indicate the best scores among the results obtained by those methods before post-processing (PP) whereas \textcolor{red}{red numbers} are those mean Dice scores under 0.700. FT: fine-tuning using the synthetic LGE dataset. N/A means that the ASD value cannot
be calculated due to missing predictions for that cardiac structure.}}
\label{tab: result table}
\begin{threeparttable}
\begin{tabular}{lllllrrrr}
\toprule
 & \multicolumn{4}{c}{\textbf{Dice}} & \multicolumn{4}{c}{\textbf{ASD}} \\
\multirow{-2}{*}{\textbf{Method}} & \multicolumn{1}{c}{LV} & \multicolumn{1}{c}{MYO} & \multicolumn{1}{c}{RV} & \multicolumn{1}{c}{AVG*} & \multicolumn{1}{c}{LV} & \multicolumn{1}{c}{MYO} & \multicolumn{1}{c}{RV} & \multicolumn{1}{c}{AVG*} \\ \midrule
MIRTK & 0.819 & {\color[HTML]{FE0000} 0.665} & 0.831 & 0.772 & 2.56 & 1.65 & 2.11 & 2.11 \\
U-net & {\color[HTML]{FE0000} 0.624} & {\color[HTML]{FE0000} 0.441} & {\color[HTML]{FE0000} 0.577} & {\color[HTML]{FE0000} 0.547} & 10.03 & 6.07 & N/A & N/A \\
% U-net (S) & 0.862 & 0.750 & 0.894 & 0.835 & 1.93 & 1.55 & 1.15 & 1.54 \\
U-net (FT) & 0.874 & 0.781 & 0.896 & 0.850 & 1.78 & 1.50 & 1.28 & 1.52 \\
Cascaded U-net (FT) & {\color[HTML]{3531FF} 0.895} & {\color[HTML]{3531FF} 0.812} & {\color[HTML]{3531FF} 0.898} & {\color[HTML]{3531FF} 0.868} & {\color[HTML]{3531FF} 1.41} & {\color[HTML]{3531FF} 1.46} & {\color[HTML]{3531FF} 1.23} & {\color[HTML]{3531FF} 1.37} \\ \hline
Cascaded U-net (FT) + PP & 0.897 & 0.816 & 0.895 & 0.869 & 1.17 & 1.42 & 1.18 & 1.26\\ \bottomrule
% \\
% Cascaded U-net (FT) + PP (EMMA) & 0.904 & 0.819 & 0.895 & 0.873 & 1.15 & 1.36 & 1.17 & 1.22 \\ \hline
\end{tabular}
\begin{tablenotes}
\item * For ease of comparison, we calculate the average (AVG) Dice score and the average ASD score over the three structures for each method. 
\end{tablenotes}
\end{threeparttable}
\end{table}

%for camera-ready% 
Finally, we applied ensemble learning to improve our model's performance in the test phase. Specifically, we trained the proposed segmentation network for multiple times, each time regenerating a new synthetic LGE dataset for fine-tuning. We trained four models in total. Our final submission result for each test image was obtained by averaging the probabilistic maps from these models and then assigning to each pixel the class with the highest score. In the testing stage of the competition, the method achieves very promising segmentation performance on a relative large test set (40 subjects), with an average Dice score of 0.92 for LV, 0.83 for MYO, and 0.88 for RV; an ASD of 1.66 for LV, 1.76 for MYO, and 2.16 for RV.

\section{Conclusion}
In this paper, we showed that synthesizing multi-modal LGE images from labelled bSSFP images to finetune a pre-trained segmentation network shows impressive segmentation performance on LGE images even though the  network has not seen \emph{real} labelled LGE images before. We also demonstrated that the proposed segmentation network (Cascaded U-net) outperformed the baseline methods by a significant margin,  suggesting the benefit of integrating the high-level shape knowledge and low-level image appearance to guide the segmentation procedure. More importantly, our cascaded segmentation network is independent of the particular architecture of underlying convolutional neural networks. In other words, the basic neural network (U-net) in our work can  be  replaced  with  any  of the state-of-the-art segmentation network to  potentially  improve  the  prediction  accuracy  and robustness. Moreover, the proposed solution based on unsupervised multi-modal style transfer is not  only limited  to the cardiac  image  segmentation but can be extended to other multi-modal image analysis tasks where the manual annotations of one modality are not available. Future work will focus on the  application of the method to the problems such as domain adaptation for multi-modality brain segmentation.
\bibliographystyle{unsrt}
\bibliography{mybib}

\begin{thebibliography}{10}

\bibitem{zhuang2016multivariate}
Xiahai Zhuang.
\newblock {Multivariate mixture model for cardiac segmentation from
  multi-sequence MRI}.
\newblock In {\em MICCAI}, pages 581--588, 2016.

\bibitem{lu2013automatic}
YingLi Lu et~al.
\newblock {Automatic myocardium segmentation of LGE MRI by deformable models
  with prior shape data}.
\newblock {\em JCMR}, 15(1):P14, 2013.

\bibitem{tao2015automated}
Qian Tao et~al.
\newblock {Automated left ventricle segmentation in late gadolinium-enhanced
  MRI for objective myocardial scar assessment}.
\newblock {\em JMRI}, 42(2):390--399, 2015.

\bibitem{Zhuang2018MultivariateMM}
Xiahai Zhuang.
\newblock {Multivariate mixture model for myocardium segmentation combining
  multi-source images}.
\newblock {\em PAMI}, 2018.

\bibitem{yue2019cardiac}
Qian Yue et~al.
\newblock {Cardiac Segmentation from LGE MRI Using Deep Neural Network
  Incorporating Shape and Spatial Priors}.
\newblock {\em MICCAI}, 2019.

\bibitem{ronneberger2015u}
Ronneberger et~al.
\newblock {U-net: Convolutional networks for biomedical image segmentation}.
\newblock In {\em MICCAI}, 2015.

\bibitem{huang2018multimodal}
Xun Huang et~al.
\newblock {Multimodal unsupervised image-to-image translation}.
\newblock In {\em ECCV}, 2018.

\bibitem{Qin_2019_IPMI}
Chen Qin, Bibo Shi, Rui Liao, Tommaso Mansi, Daniel Rueckert, and Ali Kamen.
\newblock Unsupervised deformable registration for multi-modal images via
  disentangled representations.
\newblock In {\em Information Processing in Medical Imaging}, pages 249--261.
  Springer, Cham, June 2019.

\bibitem{Bengio2009}
Yoshua Bengio, J{\'e}r{\^o}me Louradour, Ronan Collobert, and Jason Weston.
\newblock Curriculum learning.
\newblock In {\em Proceedings of the 26th Annual International Conference on
  Machine Learning}, ICML '09, pages 41--48, New York, NY, USA, 2009. ACM.

\bibitem{Tu2010}
Zhuowen Tu and Xiang Bai.
\newblock {Auto-context and its application to high-level vision tasks and 3D
  brain image segmentation}.
\newblock {\em PAMI}, 2010.

\bibitem{sobel}
Irwin Sobel and G~Feldman.
\newblock A 3x3 isotropic gradient operator for image processing.
\newblock {\em Pattern Classification and Scene Analysis}, pages 271--272, 01
  1973.

\bibitem{krahenbuhl2011efficient}
Philipp Kr{\"a}henb{\"u}hl et~al.
\newblock {Efficient inference in fully connected CRFs with gaussian edge
  potentials}.
\newblock In {\em NeuralIPS}, 2011.

\bibitem{huang2017arbitrary}
Xun Huang et~al.
\newblock {Arbitrary style transfer in real-time with adaptive instance
  normalization}.
\newblock In {\em ICCV}, 2017.

\bibitem{chen_2019}
Chen Chen, Wenjia Bai, and Daniel Rueckert.
\newblock Multi-task learning for left atrial segmentation on {GE-MRI}.
\newblock In {\em Statistical Atlases and Computational Models of the Heart.
  Atrial Segmentation and LV Quantification Challenges}, 2019.

\end{thebibliography}
\newpage
\section*{Supplemental Material}
\begin{figure}[!h]
    \centering
    \includegraphics[width=1.0\textwidth]{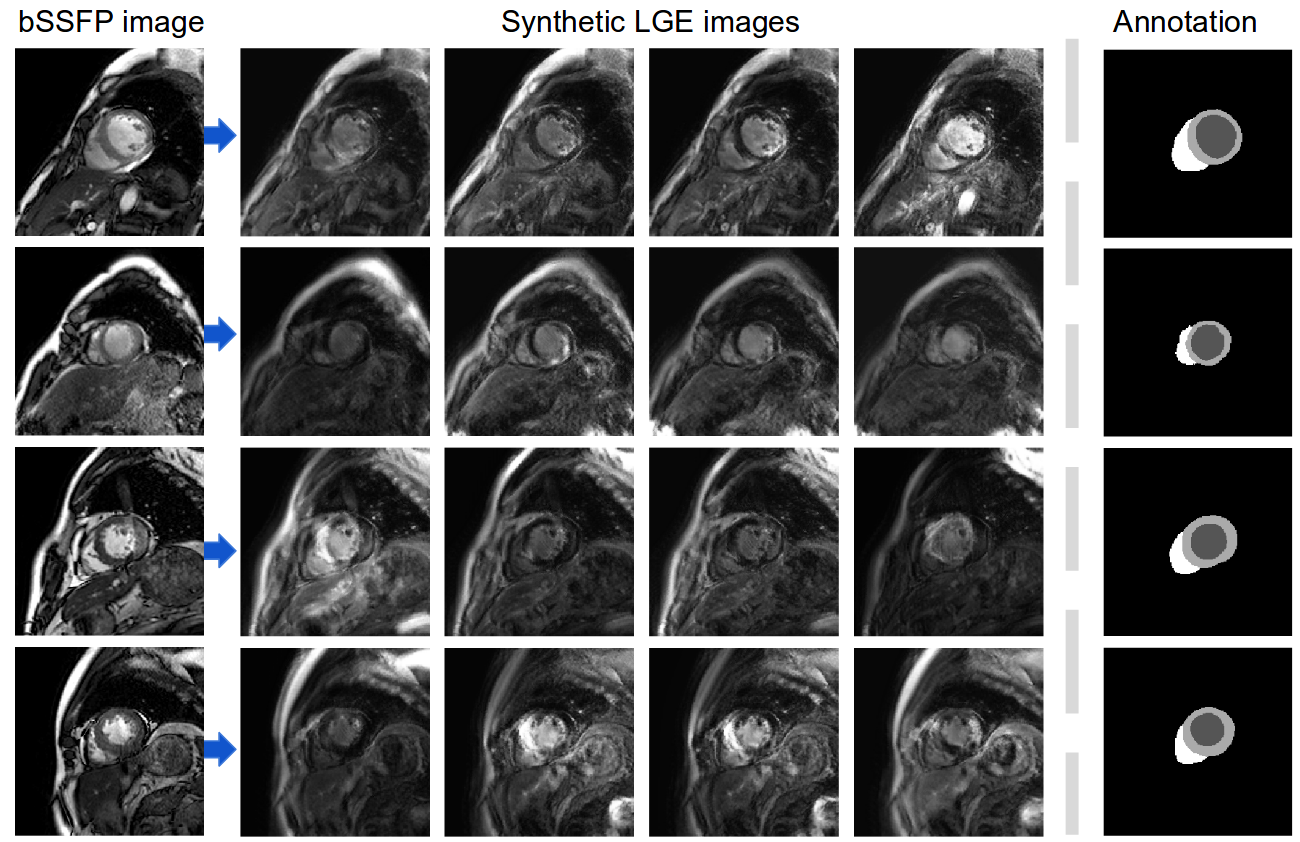}
    \caption{\textbf{Exemplar synthetic LGE images generated from bSSFP images using the multi-modal image translation network.} Given \textbf{one} bSSFP image (column 1), the translation network translates the image into \textbf{multi-modal} LGE-like images (column 2 to 4). These translated images differ in image brightness and contrast as well as the intensity distribution in the cardiac region, while preserving the same cardiac anatomy. These synthetic images, in together with the annotations on the original bSSFP images (the last column) contribute to the synthetic dataset which is used to fine-tune the proposed segmentation network. }
    \label{fig:trans_demo}
\end{figure}
\end{document}